\documentclass[aps,pra,twocolumn,showpacs,preprintnumbers,amsmath,amssymb]{revtex4-1}
\usepackage{graphicx}
\usepackage[usenames,dvipsnames]{xcolor}
\usepackage{hyperref}
\usepackage{cleveref}
\usepackage{makecell}
\usepackage{float}
\usepackage{lineno}
\usepackage[T1]{fontenc}
\usepackage[caption=false]{subfig}
\newcommand{\bra}[1]{\ensuremath{\langle #1 |}}
\newcommand{\ket}[1]{\ensuremath{| #1 \rangle}}

\usepackage{amssymb,amsmath} 
\usepackage{amsfonts}
\usepackage{color}
\usepackage[utf8]{inputenc}
\usepackage{lmodern}
\usepackage{epsfig}
\usepackage{ulem}
\usepackage{blindtext}
\begin{document}
\title{Phase diagram of Rydberg atoms in a two-leg rectangular ladder}

\author{Shu-Ao Liao}
\author{Jin Zhang}
\email{jzhang91@cqu.edu.cn}
\author{Li-Ping Yang}
\email{liping2012@cqu.edu.cn}
\affiliation{Department of Physics and Chongqing Key Laboratory for Strongly Coupled Physics, Chongqing University, Chongqing 401331, China}
\definecolor{burnt}{cmyk}{0.2,0.8,1,0}
\def\lt{\lambda ^t}
\def\note{note}
\def\beq{\begin{equation}}
\def\enq{\end{equation}}

\date{\today}
\begin{abstract}
Using the density matrix renormalization group algorithm, we map the ground-state phase diagram of a two-leg Rydberg ladder array with lattice spacings $a_x=2a_y$. We identify various density wave phases that spontaneously break the translational symmetry or the top-bottom reflection symmetry within the ladder. By increasing the laser detuning from zero, where the system is in a disordered phase that preserves all symmetries, we observe density wave orders with spontaneous breaking of the translational $\mathbb{Z}_p$ symmetries at intermediate detuning values, while the reflection symmetry is preserved. These orders exhibit nonzero bond orders with positive expectation values on every $p$th rung, thus labeled as $\mathbb{Z}_p^+$ phases. At larger detuning values, another spontaneous breaking of the reflection symmetry, which disrupted the bond orders on the rungs, occurs via an Ising phase transition. In these phases, either the top or the bottom site is occupied in a staggered way on every $p$th rung, breaking the translational $\mathbb{Z}_{2p}$ symmetry, thus labeled by $\mathbb{Z}_{2p}$ phases. We locate and characterize the 3-state Potts point and Ashkin-Teller point along the commensurate lines, as well as the direct chiral phase transitions between the disordered phase and the $\mathbb{Z}_p^+$ ($p = 3, 4$) phases. Critical exponents $\nu$ and $z$ are calculated for both conformal and chiral phase transition points. We finally identify two types of floating phases in the phase diagram: one characterized by a quasi-long-range incommensurate bond-order wave, and the other by a quasi-long-range incommensurate wave of density differences in the rungs. Our work motivates further applications of Rydberg atom arrays in quantum simulation.
\end{abstract}


\maketitle

\section{Introduction}\label{sec:introduction}

The study of strongly correlated quantum systems has gained significant attention due to their potential to reveal novel phases of matter and exotic quantum criticalities. Among the various techniques for addressing such problems, quantum simulation has proven to be a highly promising method for providing answers to open questions that cannot be solved with classical computations. Recently, Rydberg atoms, which have long lifetimes and large electric dipole moments, have emerged as a highly favorable platform for simulating diverse strongly correlated quantum many-body physics \cite{Labuhn2016RydIsing,Bernien2017Dynamics,Keesling2019Kibble,Leseleuc2019topo,Semeghini2021SL,Ebadi2021_256,Pascal2021AF,ChenContinuous2023,Daniel2024RydStringBreaking}. In these quantum simulation schemes, neutral atoms in their ground states are trapped by optical tweezers to create a defect-free atom array, while high-lying excited Rydberg states are coupled to the ground states by lasers. With van der Waals interactions between Rydberg atoms and laser detuning serving as a chemical potential, the effective model Hamiltonian becomes an Ising model with density-density interactions that decay as $1/r^6$.

Critical phenomena can be classified into different universality classes that depend solely on symmetries and dimensionality, thereby providing unifying principles applicable across various fields of physics. In Rydberg systems, within a characteristic distance known as the Rydberg blockade radius, at most one atom can be excited to the Rydberg state \cite{Jakschfastgate2000,LukinDipoleBlockade2001,DudinRabi2012,RydPhysics2018,BrowaeysIndividual2020}. 
This blockade mechanism, combined with the configurable geometry of the tweezer array, makes the Rydberg quantum simulator highly programmable and capable of simulating quantum criticalities associated with the spontaneous breaking of different symmetries in various dimensions. Consequently, two-dimensional (2D) and three-dimensional (3D) Ising-type phase transitions have been experimentally probed using one-dimensional (1D) Rydberg-atom chains \cite{Bernien2017Dynamics} and 2D Rydberg square arrays \cite{Ebadi2021_256,Pascal2021AF}, respectively. Other types of phase transitions, such as Potts, Ashkin-Teller (AT), Chiral, Berezinskii–Kosterlitz–Thouless (BKT), and Pokrovsky-Talapov (PT) transitions in $1+1$ dimensions \cite{PhysRevLett.122.017205,PhysRevResearch.3.023049,PhysRevResearch.4.043102}, and Potts and O($\mathcal{N}$) universality classes in $2+1$ dimensions \cite{PhysRevLett.124.103601,PhysRevLett.130.206501,RhineKagome2021,CXLiTriangular2022}, have been predicted by numerical studies. Additionally, the Kibble-Zurek mechanism \cite{Keesling2019Kibble,Chepiga&Mila2021Kibble}, strongly correlated spin transport \cite{PhysRevX.14.011025}, emergent glassy behaviors \cite{PhysRevLett.130.206501}, and exotic quantum phases \cite{rader2019floating,zhang2024probingquantumfloatingphases,PhysRevB.106.115122,PhysRevLett.131.203003,PhysRevLett.124.103601,PhysRevE.106.034121,Semeghini2021SL,Samajdar:2022mtt,PhysRevX.11.031005,PhysRevX.12.041029,PhysRevLett.132.206503,TrimerLiquidQuEra2023,PhysRevA.108.053314} have been explored in Rydberg atom arrays theoretically or experimentally.

Although the phase diagram of the 1D Rydberg chain is well understood \cite{rader2019floating,PhysRevResearch.4.043102}, and various quantum phenomena in 2D Rydberg arrays with different geometries have been explored \cite{Ebadi2021_256,Pascal2021AF,Semeghini2021SL,ChenContinuous2023,PhysRevLett.124.103601,PhysRevLett.130.206501,RhineKagome2021,CXLiTriangular2022,Samajdar:2022mtt,PhysRevX.11.031005,PhysRevX.12.041029,PhysRevLett.132.206503,TrimerLiquidQuEra2023,PhysRevA.108.053314}, quasi-1D Rydberg arrays with interacting multiple chains remain less investigated. In two-leg square and triangular ladders with particle-conserving Rydberg-dressed atoms, by selecting a filling factor that supports clustering in the classical limit, rich quantum phases and quantum criticalities with different central charges have been identified in the phase diagram \cite{PhysRevB.105.155159,Fromholz:2022ymy}. In Rydberg tweezer ladders, where there is no U(1) charge conservation, quantum order-by-disorder-induced Ising phase transitions have been discussed in the small blockade radius regime \cite{SarkarOrderbyDis2023,JinPRDCritical2024}. In the square ladder with a larger blockade radius that allows at most one Rydberg state in each square, $\mathbb{Z}_3^{\pm}$ orders with rung bond-order density waves are observed, where the ladder's top-bottom reflection symmetry remains unbroken \cite{eck2023critical}. However, these studies cut off the van der Waals interactions at small distances, and the physics in the stronger interaction regime remains an open question. Recently, using QuEra's Rydberg quantum simulator, a two-leg Rydberg rectangular ladder with lattice spacings $a_y=2a_x$ was constructed, and the floating phase was experimentally probed \cite{zhang2024probingquantumfloatingphases}. The phase diagram was mapped out numerically by considering all interactions within any consecutive 20 rungs, revealing that all crystalline orders for $R_b/a_x < 3.5$ break the top-bottom reflection symmetry, and phases like $\mathbb{Z}_3^{\pm}$ are absent. This suggests that the aspect ratio of the ladder plays a crucial role in determining the phase diagram. Given these complexities in Rydberg ladder systems, our study seeks to further elucidate the phase behavior under varying interaction strengths and lattice configurations.

In this paper, we study the phase diagram of a Rydberg rectangular ladder with $a_x = 2a_y$ using the density matrix renormalization group (DMRG) algorithm \cite{PhysRevLett.69.2863,PhysRevB.48.10345}. We identify various density wave orders that spontaneously break translational symmetry, top-bottom reflection symmetry, or both. By increasing the laser detuning in the disordered phase, the translational $\mathbb{Z}_p$ symmetry is broken first while the reflection symmetry remains preserved, leading to the formation of $\mathbb{Z}_p^+$ orders characterized by a rung bond-order wave. 
Further increasing the detuning results in the breaking of reflection symmetry via an Ising transition, leading to the $\mathbb{Z}_{2p}$ phase, where the period of the density wave doubles. We characterize the Ising, Potts, and AT conformal field theory (CFT) phase transitions, as well as Huse \& Fisher’s chiral phase transitions \cite{HuseFisherPRL1982}. The critical correlation length exponent $\nu$ and dynamical exponent $z$ are calculated for the chiral transitions near the CFT points, confirming that these CFT points correspond to extremal values of $\nu$ and $z$ as functions of the parameters. In the strong interaction regime, we find two types of floating phases between crystalline orders: one characterized by a quasi-long-range (QLR) incommensurate bond-order wave and the other by a QLR incommensurate wave in the Rydberg density differences across the rungs. The transitions out of the floating phase into the disordered phase and crystalline orders belong to the BKT and PT universality classes, respectively.

This paper is organized as follows. In Section \ref{sec:model}, we introduce the Hamiltonian that describes the Rydberg ladder system and discuss the roles of its parameters as well as the possible phases in the ground-state phase diagram. We then outline the key quantities used to characterize quantum phases and phase transitions and specify the parameter settings employed in our DMRG algorithms. The phase diagram and an overview of all identified phases and phase transitions are presented in Sec.~\ref{subsec:phasediagram}. A detailed analysis of the disordered phase is provided in Sec.~\ref{subsec:disorder}, followed by investigations of the Ising transitions in Sec.~\ref{subsec:ising}, the Potts CFT and chiral transitions for the $\mathbb{Z}_3^+$ order in Secs.~\ref{subsec:z3pchiral}, the AT CFT and chiral transitions for the $\mathbb{Z}_4^+$ order in Sec.~\ref{subsec:z4pchiral}, and the floating phase in Sec.~\ref{subsec:floating}. Finally, we summarize our results and discuss directions for future work in Sec.~\ref{sec:conclusion}.

\section{Model and Methods}\label{sec:model}

\subsection{Model Hamiltonian} \label{subsec:qo2model}
We study a two-leg Rydberg ladder array with lattice spacings $a_x = 2 a_y = a$, where $a_x$ is the distance between nearest-neighbor (NN) rungs and $a_y$ is the spacing between the two legs. In this configuration, the interaction between two Rydberg atoms within the same rung is significantly stronger than the interaction between atoms in different rungs. The effective Hamiltonian of the system is given by
\begin{eqnarray}\label{eq:rydbergham}
\nonumber \hat{H} &=& \sum_{i=1}^{L} \sum_{s = 1}^{2} \left( \frac{\Omega}{2}  \ket{g_{i,s}}\bra{r_{i,s}} + \text{h.c.} - \Delta \hat{n}_{i,s} \right)  \\
&+&  \sum_{\mathbf{r} \neq {\mathbf{r’}}} V_{\mathbf{r}, \mathbf{r’}} \hat{n}_{\mathbf{r}}\hat{n}_{\mathbf{r’}},
\end{eqnarray}
where $i$ labels the rungs and $s = 1, 2$ labels the legs. In the experiment, each atom, initially in the ground state $\ket{g}$, is trapped in optical tweezers and then excited to the Rydberg state $\ket{r}$ with a large principal quantum number via laser excitation. The Rabi frequency $\Omega$ is effectively generated by a two-photon process, $\Delta$ represents the laser detuning, and $V_{\mathbf{r}, \mathbf{r}’}$ denotes the interaction between Rydberg atoms at positions $\mathbf{r}$ and $\mathbf{r}’$. The position vector $\mathbf{r}$ in the two-leg system is given by $\mathbf{r} = i a_x \mathbf{e}_x + s a_y \mathbf{e}_y$. The Rydberg state density operator is $\hat{n}_i = \ket{r_i}\bra{r_i}$. The interaction between Rydberg atoms follows a van der Waals form, $V_{\mathbf{r}, \mathbf{r}’} = C_6 / |\mathbf{r} - \mathbf{r}’|^6$, where $C_6$ is a constant. We parametrize the Hamiltonian using the Rydberg blockade radius $R_b$, defined by $C_6 / R_b^6 = \Omega$, where the interaction at distance $R_b$ is equal to the Rabi frequency. In practical calculations, we retain interactions within any set of consecutive twenty-one rungs to simulate experimental data. We also set $\Omega = 1$ and vary $R_b$ in units of $a_x$, i.e., $R_b / a$, as well as the detuning $\Delta / \Omega$, to map out the phase diagram.

The detuning $\Delta$ acts as a chemical potential in the effective Hamiltonian, with the total number of Rydberg excitations increasing as $\Delta$ is raised. However, the strong repulsive interaction between Rydberg atoms restricts the minimum distance between them, limiting the overall Rydberg density. The competition between these two effects gives rise to various translational symmetry-breaking phases characterized by density waves, where every $p$th site in each leg of the two-leg system is occupied by Rydberg atoms. In the absence of top-bottom $\mathbb{Z}_2$ symmetry breaking, double occupation of Rydberg atoms in the same rung is forbidden for large $R_b$, resulting in entangled states of the form $(\ket{g_{i,1}}\ket{r_{i,2}} \pm \ket{r_{i,1}}\ket{g_{i,2}})/\sqrt{2}$ in the occupied rungs \cite{LukinDipoleBlockade2001}. This bond-order density wave with an integer period $p$ is referred to as the $\mathbb{Z}_p^{\pm}$ phase. We will demonstrate that only the $\mathbb{Z}_p^{+}$ phase appears in our phase diagram. When top-bottom $\mathbb{Z}_2$ symmetry is broken, the Rydberg atoms in the top leg are shifted to reside on the perpendicular bisector of the NN occupied sites on the bottom leg to minimize interaction energy, doubling the translational period of the density waves. These phases are labeled as $\mathbb{Z}_{2p}$ phases.

By increasing $\Omega$, strong quantum fluctuations will melt the $\mathbb{Z}_p^{+}$ and $\mathbb{Z}_{2p}$ crystalline orders into the disordered phase. Previous studies on the Rydberg chain \cite{PhysRevLett.122.017205,Chepiga&Mila2021Kibble,PhysRevResearch.4.043102} have shown that the melting transition for commensurate density waves with periodicity 2 is in the Ising universality class. For commensurate density waves with periodicities of 3 or 4, the transition can be a CFT point in the Potts or AT universality class at specific parameter values, non-CFT points in the chiral universality class \cite{HuseFisherPRL1982}, or involve two-step transitions through a quasi-long-range ordered phase called the floating phase. For commensurate density waves with periodicity 5 or higher, there is no direct transition into the disordered phase and a floating phase always intervenes. Since the floating phase lies outside and between commensurate crystalline orders, it is an incommensurate phase with algebraically decaying correlation functions. We will demonstrate that these phase transitions are also observed around the $\mathbb{Z}_p^{+}$ phases in our system. Given the two types of crystalline orders, with or without bond-order waves, there are also two types of floating phases, characterized by the presence or absence of quasi-long-range bond-order waves. The floating phase without quasi-long-range bond-order waves can be characterized by a quasi-long-range wave order of density differences of the two sites in the rungs.
\subsection{Entanglement entropy}
\label{sec:ent}
One universal tool to detect quantum phase transitions is the von Neumann entanglement entropy $\mathcal{S}_{\mathrm{vN}}$. For a quantum many-body system divided into two parts, $\mathcal{A}$ and $\mathcal{B}$, the bipartite entanglement between $\mathcal{A}$ and $\mathcal{B}$, assuming the system is in the ground state $\ket{\Psi_0}$, is characterized by $\mathcal{S}_{\mathrm{vN}} = -\operatorname{Tr} \rho_{\mathcal{A}} \ln \rho_{\mathcal{A}}$, 
where $\rho_{\mathcal{A}} = \operatorname{Tr}_{\mathcal{B}}\left(\left|\Psi_0\right\rangle\left\langle\Psi_0\right|\right)$ is the reduced density operator of subsystem $\mathcal{A}$. In gapped quantum phases, the scaling of the entanglement entropy follows an area law, while it may have logarithmic corrections in critical systems \cite{RevModPhys.82.277}. For 1D quantum systems with open boundary conditions (OBC), CFT predicts that the entanglement entropy of a critical point has the following form \cite{AffleckCritical1991,HOLZHEY1994443,VidalEntangle2003,PasqualeCalabrese_2004,Calabrese_2009}:
\begin{eqnarray}
\label{eq:cfteeform}
\mathcal{S}_{\mathrm{vN}}= \frac{c}{6} \ln \left\{\frac{4\left(L+1\right)}{\pi} \sin \left[\frac{\pi\left(2 l+1\right)}{2\left(L+1\right)}\right]\right\}+\ln{g} + s_o,
\end{eqnarray}
where $c$ is the central charge with universal values for certain types of CFT phase transitions, $\ln{g}$ is the boundary entropy associated with the ground state degeneracy, $s_{o}$ is a non-universal constant, and $L$ and $l$ are the sizes of the entire system and subsystem $\mathcal{A}$, respectively. In 1D systems, the CFT form of $\mathcal{S}_{\rm{vN}}$ in Eq.~\eqref{eq:cfteeform} typically applies to continuous phase transition lines associated with spontaneous symmetry breaking and gapless phases described by Tomonaga–Luttinger liquid theory \cite{F.D.M.Haldane_1981,PhysRevB.84.085114}, up to some oscillatory terms that vanish for large $L$ with fixed $l/L$ \cite{LaflorencieBoundaryEE2006,CalabreseParityprl2010,XavierRenyiParity2011,BazavovEstimating2017,UnmuthProbingPRA2017}. For a fixed ratio $l / L$, $\mathcal{S}_{\rm{vN}}$ at CFT points diverges logarithmically with $L$. Different scalings hold for non-CFT critical points, where the divergence of $\mathcal{S}_{\rm{vN}}$ can be faster than the CFT form \cite{TempleHeEELifshitz2017}. These properties can be used to identify critical lines and critical phases in the ground-state phase diagram of 1D quantum systems.

\subsection{Structure Factor}
The structure factor is commonly used in experimental physics to detect periodic crystalline orders and can be directly measured in scattering experiments. It is the Fourier transform of the correlation function and serves as a theoretical tool for characterizing true or quasi-long-range orders. Given the two types of crystalline orders — those with and without a bond-order wave — we employ two distinct structure factors to characterize the quantum phases of our system:
\begin{eqnarray}
\label{eq:strucm}
S_{m}(k) &=& \frac{p_o^2}{4L^2} \sum_{i,i’} e^{\mathrm{i} k (i-i’)}\langle \hat{m}_{i} \hat{m}_{i’} \rangle, \\
\label{eq:strucb}
S_{B}(k) &=& \frac{p_o^2}{L^2} \sum_{i,i’} e^{\mathrm{i} k (i-i’)}\langle \hat{B}_{i} \hat{B}_{i’}\rangle,
\end{eqnarray}
where $\hat{m}_i = \hat{n}_{i,2} - \hat{n}_{i,1}$ is the Rydberg density difference between the two sites in the $i$th rung, and $\hat{B}_i = \hat{a}^{\dagger}_{i,1}\hat{a}_{i,2}+\hat{a}^{\dagger}_{i,2}\hat{a}_{i,1}$ is the bond-order operator. Here, $p_o = 2\pi/k_o$, with $k_o$ being the peak position of the structure factor.

In the classical limit, the $\mathbb{Z}_{2p}$ phase has $\langle \hat{m}_i \rangle$ taking values of $1$ and $-1$ at equidistant intervals with zeros in between, resulting in a unit cell size of $p_o = 2p$. In Eq. \eqref{eq:strucm} at $k=k_o$, there are $(L/p)^2$ instances of ones in the sum, so the normalization factor should be $(2L/p_o)^2$. The $\mathbb{Z}^+_p$ phase has $\langle \hat{B}_i\rangle$ equal to $1$ in every $p$th rung, with all other values zero. Therefore, there are $(L/p_o)^2$ non-zero terms in the sum in Eq.~\eqref{eq:strucb} at $k=k_o$, and the normalization factor should be $(L/p_o)^2$.

For true long-range orders with commensurate density waves (crystalline orders), where periodicity is an integer multiple of the lattice spacing, the structure factor peaks at $2\pi/p$ and $2\pi-2\pi/p$, with $p$ being an integer. In 1D quantum systems, incommensurate phases with non-integer multiples of the lattice spacing can manifest as quasi-long-range orders (QLROs) with correlation functions decaying algebraically as $\sim 1/r^\eta$. In the disordered phase, correlation functions also exhibit commensurate or incommensurate oscillations, with amplitude decaying exponentially with a correlation length $\xi$ \cite{PhysRevResearch.4.043102,zhang2024probingquantumfloatingphases}. As demonstrated in Appendix \ref{apsec:strucpeak}, for $\eta \le 2$ and moderately large $\xi$, the structure factor will peak at the wave vector of QLROs and can accurately indicate the wave vector of the disordered phase \cite{zhang2024probingquantumfloatingphases}.

\subsection{Binder Cumulant}

The continuous phase transition from crystalline orders to the disordered phase can be probed using order parameters and their higher-order cumulants. A widely used quantity to study phase transitions is the Binder cumulant \cite{BINDER19851}:
\begin{eqnarray}
\label{eq:bindercumulant}
U_4 = 1-\frac{\langle \hat{M}^{4}_{k}\rangle}{3\langle \hat{M}^{2}_{k}\rangle^{2}},
\end{eqnarray}
where $\hat{M}_k$ can be $\hat{M}_k^m = (\sum_i \cos\left(ki\right)\hat{m}_i)/L$ and $\hat{M}_k^B = (\sum_i \cos\left(ki\right)\hat{B}_i)/L$ to  represent the order parameters for the $\mathbb{Z}_{2p}$ and $\mathbb{Z}_{p}^+$ phases, respectively, with density waves at wave vector $k$.

Approaching the phase transition point at fixed $R_b/a$, the correlation length diverges as $\xi \sim |\Delta / \Omega-(\Delta / \Omega)_c |^{-\nu}$, where $(\Delta / \Omega)_c$ is the critical point and $\nu$ is the correlation length exponent. Since $U_4$ is dimensionless, the critical point is identified as a fixed point of $U_4$, which manifests as a crossing point for $U_4$ curves plotted against $\Delta/\Omega$ for different system sizes. Moreover, the scaling hypothesis suggests that near the critical point, $U_4$ follows the functional form
\begin{equation}
\label{eq:U4universalfunction}
U_4 = f\left[ L^{1/\nu} (\Delta/\Omega - (\Delta/\Omega)_{c}) \right],
\end{equation}
up to subleading corrections that vanish in the thermodynamic limit, where $f(x)$ is a universal function. In practice, we calculate values of $U_4$ around the crossing point for different $L$s and fit the $U_4$ data to a high-degree polynomial of $L^{1/\nu} (\Delta/\Omega - (\Delta/\Omega)_{c})$. By tuning $(\Delta/\Omega)_{c}$ and $\nu$, we can achieve the best data collapse by minimizing the sum of squares of residuals, which determines the values of the critical point and the correlation length exponent.

\subsection{Gap scaling}
At a critical point, the gapless spectrum of a quantum system has a dispersion relation $\omega \sim k^z$, where $z$ is the dynamical exponent. This implies that the energy gap for finite systems scales as $\Delta E \sim L^{-z}$, and the characteristic timescale of the system near the continuous phase transition point follows $\tau_{c} \sim \xi^{z}$ \cite{HohenbergDynamic1977}. For instance, when $z=1$, the phase transition point is described by CFTs, whereas $z>1$ indicates a non-CFT transition, such as those in the chiral universality class. The value of $z$ can be determined using the scaling form of the energy gap near the critical point:
\begin{eqnarray}
\label{eq:dEuniversalfunction}
L^z \Delta E = g\left[L^{1 / \nu} \left(\Delta / \Omega - (\Delta / \Omega)_c\right)\right],
\end{eqnarray}
where $g(x)$ is a universal function different from $f(x)$ in Eq.~\eqref{eq:U4universalfunction}, and $\Delta E$ is the energy gap between the ground state and the first excited state. The data collapse procedure for $\Delta E$ follows the same method as that used for $U_4$.

\subsection{Parameters of DMRG algorithms}

We perform finite-size DMRG calculations based on matrix product states (MPS) \cite{PhysRevLett.75.3537} to determine the ground states of our system. The code is implemented using the \textsc{ITensor Julia Library} \cite{10.21468/SciPostPhysCodeb.4}, retaining all Rydberg interactions within any twenty-one consecutive rungs in the Hamiltonian to closely simulate experimental conditions. For ground-state searches, we gradually increase the maximal bond dimension $D$ during the variational sweeps until the truncation error $\epsilon$ falls below $10^{-10}$. Calculations using different truncation errors are specified where relevant. DMRG sweeps are terminated once the change in ground-state energy is less than $10^{-11}$ and the von Neumann entanglement entropy changes less than $10^{-8}$ in the final two sweeps. System sizes are selected to match the crystalline orders. Generally, tens of sweeps are enough for convergence in crystalline orders, while the quantum floating phase typically requires thousands of DMRG sweeps, with even more necessary for larger Rydberg blockade radius $R_b/a$.

\section{Results} \label{sec:results}

\begin{figure*}[t]
\centering 
\includegraphics[width=1\textwidth]{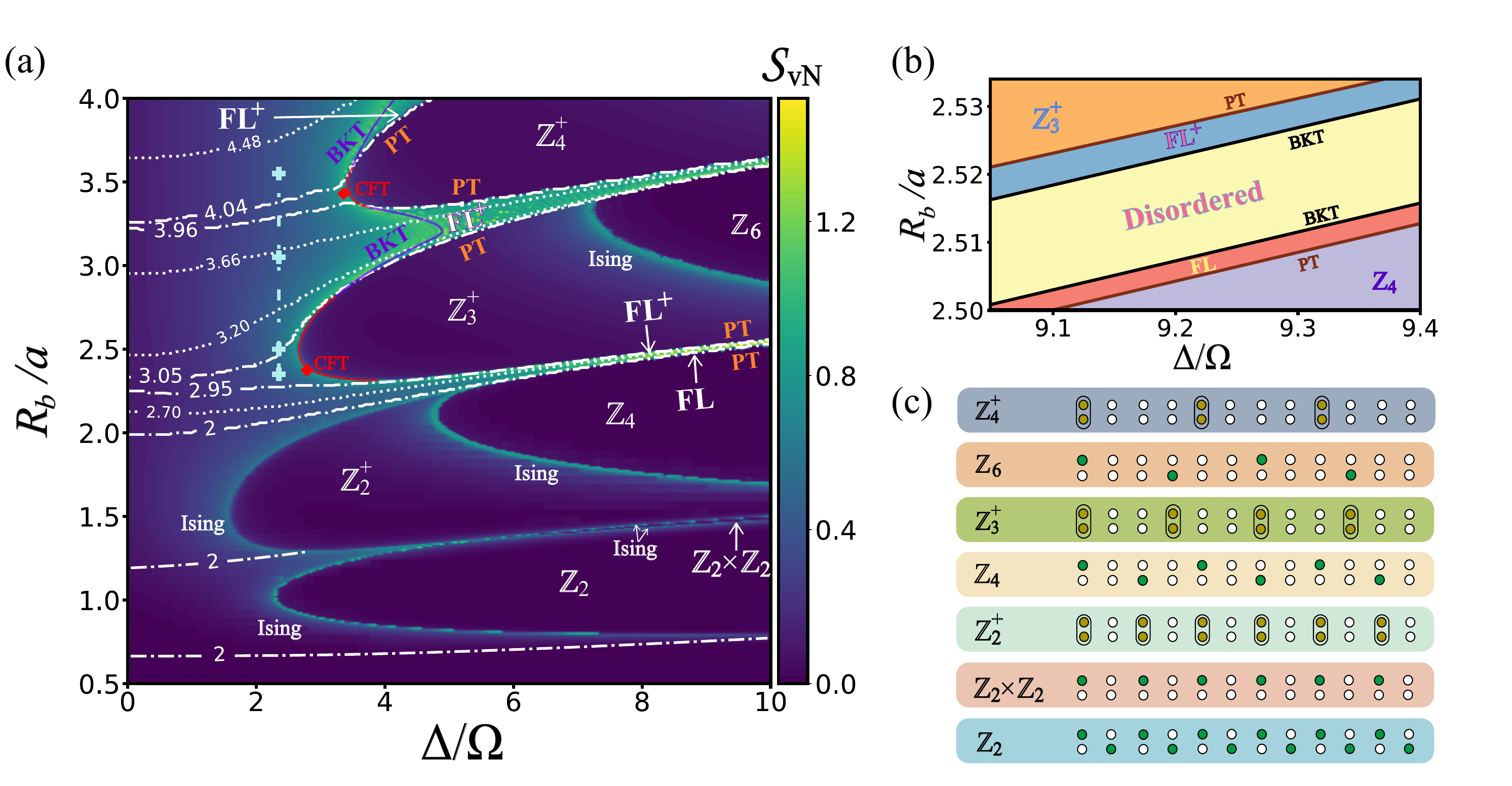}
\caption{(a) The ground-state phase diagram of the two-leg Rydberg ladder with $a_x = 2a_y$, constructed using the von Neumann entanglement entropy $\mathcal{S}_{\rm{vN}}$. The results are for $L = 289$ rungs with open boundary conditions, and $\mathcal{S}_{\rm{vN}}$ is calculated at the cut between the $144$th and $145$th rungs. The dark lobes represent crystalline orders labeled as $\mathbb{Z}_{p}$ and $\mathbb{Z}_{p}^{+}$, where $p$ indicates the periodicity of translations. The $\mathbb{Z}_2 \times \mathbb{Z}_2$ phase exhibits a Rydberg density wave in one leg only, unlike other phases, which have density waves in both legs. Two types of floating phases are present: one with an incommensurate quasi-long-range bond-order wave (FL$^+$) and the other with an incommensurate wave of density differences across the rungs (FL). Dash-dotted and dotted lines indicate constant-$p$ lines for the disordered phase with a short-range period-$p$ oscillation in the density-density correlations. Phase transitions include those of Ising universality class between the $\mathbb{Z}_p^+$ and $\mathbb{Z}_{2p}$ phases, BKT universality class between the floating and disordered phases, and PT universality class between crystalline orders and the floating phases. The $\mathbb{Z}_3^+$, $\mathbb{Z}_4^+$ and $\mathbb{Z}_4$ phases can transition directly into the disordered phase via either a CFT point or a chiral transition line. Transitions among the $\mathbb{Z}_2$, $\mathbb{Z}_2 \times \mathbb{Z}_2$, $\mathbb{Z}_2^+$, and disordered phases are all of Ising universality class. (b) A zoomed-in view of the phase diagram between the $\mathbb{Z}_4$ and $\mathbb{Z}_3^+$ phases. By increasing the blockade radius, the $\mathbb{Z}_4$ phase transitions successively into the Floating phase, the disordered phase, the Floating$^+$ phase, and finally the $\mathbb{Z}_3^+$ phase. (c) Density maps of crystalline orders in the phase diagram. Green filled circles represent atoms with high Rydberg density (close to 1), while empty circles represent atoms with low Rydberg density (close to 0). Orange filled circles in the circled rungs indicate bond orders with entangled states of the form $(\ket{r,g} + \ket{g,r}) / \sqrt{2}$.}  
\label{fig:phasediagram}
\end{figure*}

We discuss the phase diagram and the quantum phase transitions in our Rydberg ladder system in this section. Notice that the phase diagram of the Rydberg ladder with aspect ratio $a_y = 2a_x$ has been studied in Ref.~\cite{zhang2024probingquantumfloatingphases}. The lower part of the phase diagram for our system with $a_x = 2a_y$ has also been mapped out using an effective Hamiltonian in Ref.~\cite{JinPRDCritical2024}. We will compare our results with those previous studies during the following discussions.

\subsection{Phase diagram}
\label{subsec:phasediagram}

\begin{figure*}[t]
\centering 
\includegraphics[width=1\textwidth]{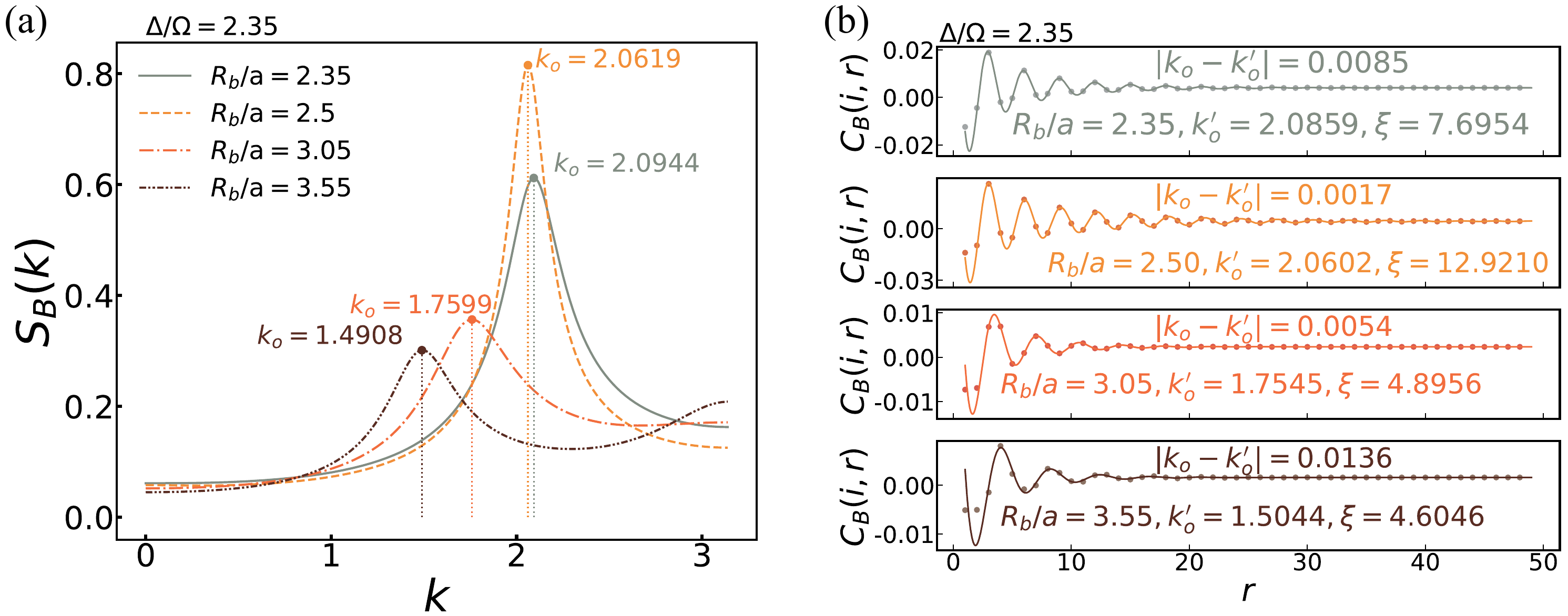}
\caption{(a) The wave vector $k_o$ of the oscillations in the correlation function $C_B(i, r) = \langle \hat{B}_i \hat{B}_{i+r} \rangle$ with $i = 143$ is determined by identifying the peak of the structure factor. (b) The wave vector $k’_o$ is extracted by fitting the correlation function to the Ornstein–Zernike form $\cos(k'_o r + \phi_0) \exp(-r/\xi)/\sqrt{r}$. The fitted correlation length $\xi$ and wave vector $k’_o$ are displayed in each subplot.}
\label{fig:DisorderStrucFit}
\end{figure*}

The phase diagram of our system is presented in Fig.~\ref{fig:phasediagram}(a). The color depth represents the magnitude of the entanglement entropy at the cut between the $144$th and $145$th rungs in a ladder of $289$ rungs. For crystalline orders, density fluctuations between different rungs have weak correlations, resulting in low entanglement entropy. One can observe various lobes of low-entanglement regimes in the phase diagram, with high-entanglement boundaries separating them from the small $\Delta/\Omega$ regime. When $\Delta = 0$, only repulsive interactions and Rabi terms are present, and Rydberg states are not favorable for occupying an extensive number of sites to form crystalline orders. Thus, the small $\Delta/\Omega$ regime corresponds to the disordered phase. We show the density maps of typical states in the low-entanglement regimes in Fig.~\ref{fig:phasediagram}(c), confirming that these regions correspond to crystalline orders. Additionally, high-entanglement lines inside the lobes indicate phase boundaries separating different crystalline orders.

There are two types of crystalline orders in our phase diagram. One type, labeled as the $\mathbb{Z}_{p}$ or $\mathbb{Z}_2 \times \mathbb{Z}_2$ phases, breaks the top-bottom reflection symmetry of the ladder, while the other type, labeled as the $\mathbb{Z}_{p}^{+}$ phase, forms an entangled state (bond order) on every $p$th rung without breaking the top-bottom reflection symmetry (see Fig.~\ref{fig:phasediagram}(c)). Here, $p$ indicates the period of translations and the degeneracy of the states. Notice that there are only $\mathbb{Z}_p$ orders with even $p$ in our system, while $\mathbb{Z}_p$ orders with odd $p$ require fine-tuning the aspect ratio of the ladder \cite{zhang2024probingquantumfloatingphases}. In the $\mathbb{Z}_p$ phase, each leg exhibits a Rydberg density wave with period $p$, but the locations of atoms with high Rydberg density $\langle \hat{n}_i \rangle$ on the two legs are shifted by $p/2$ lattice spacings to minimize the interaction energy. As a result, both translational symmetry and top-bottom reflection symmetry are broken in the $\mathbb{Z}_p$ phases. However, since flipping the ladder upside down is equivalent to a translation by $p/2$ lattice spacings, the quantum state in the $\mathbb{Z}_p$ phase keeps a degeneracy of $p$. A special phase labeled $\mathbb{Z}_2 \times \mathbb{Z}_2$ exhibits a Rydberg density wave on only one leg, resulting in a degeneracy of 4 under translations and reflections. The $\mathbb{Z}_p$ and $\mathbb{Z}_2 \times \mathbb{Z}_2$ phases can be characterized by the wave of density difference between the legs, $\hat{m}_i = \hat{n}_{i,2} - \hat{n}_{i,1}$. In the $\mathbb{Z}_p^+$ phase, the Rydberg density waves in both legs are aligned, preserving the top-bottom reflection symmetry. Due to the Rydberg blockade mechanism, double occupation of Rydberg states in a rung is forbidden, and the rungs with high Rydberg density form entangled states with a nonzero expectation value of the rung bond operator, $\hat{B}_i = \hat{a}_{i,1}^\dagger \hat{a}_{i,2} + \hat{a}_{i,2}^\dagger \hat{a}_{i,1}$. Therefore, the $\mathbb{Z}_p^+$ phase can be characterized by a bond-order wave.

The high-entanglement regimes (excluding the boundaries of crystalline orders) in the phase diagram correspond to a critical floating phase described by Luttinger Liquid theory. Similar to the single-chain case \cite{rader2019floating}, this phase exhibits QLRO with an incommensurate wave vector that varies continuously with system parameters. However, the ladder geometry introduces additional features. Between the $\mathbb{Z}_4$ and $\mathbb{Z}_3^+$ phases, there are two distinct types of floating phases: one with an incommensurate QLR bond-order wave (FL$^+$) and another with an incommensurate QLR wave of density differences in the rungs (FL) (see the zoomed-in phase diagram in Fig.~\ref{fig:phasediagram}(b)). The two floating phases are characterized by correlation functions $C_B(i, r) = \langle \hat{B}_i \hat{B}_{i+r} \rangle$ and $C_m(i, r) = \langle \hat{m}_i \hat{m}_{i+r} \rangle$, respectively, which display incommensurate oscillations with wavelengths that are not integer multiples of the lattice spacing $a_x$. Between the two floating phases lies the disordered phase, which features an incommensurate short-range density wave that decays exponentially. In our numerical results, only the FL$^+$ phase is observed between the $\mathbb{Z}_4^+$ and $\mathbb{Z}_3^+$ phases, as well as between the $\mathbb{Z}_4^+$ and $\mathbb{Z}_6$ phases, within the parameter regime considered. It is believed that crystalline orders with a period of 5 or greater should always undergo a two-step melting process via a floating phase before transitioning into the disordered phase. Therefore, there should be an FL phase above and near the $\mathbb{Z}_6$ phase in our phase diagram. However, this phase may be too narrow to identify within our parameter regime and could become more apparent at $\Delta/\Omega > 10$. Investigating this regime requires further calculations and is beyond the scope of this work.

The quantum phase transitions can be summarized as follows. Ising phase transitions occur wherever there is a spontaneous $\mathbb{Z}_2$ symmetry breaking. These include transitions between the disordered phase and the $\mathbb{Z}_2$ phase, the disordered phase and the $\mathbb{Z}_2^+$ phase, the $\mathbb{Z}_2 \times \mathbb{Z}_2$ phase and the $\mathbb{Z}_2$ phase, the $\mathbb{Z}_2^+$ phase and the $\mathbb{Z}_2 \times \mathbb{Z}_2$ phase, as well as the $\mathbb{Z}_p^+$ phase and the $\mathbb{Z}_{2p}$ phase. The $\mathbb{Z}_3^+$ phase can transition directly into the disordered phase through a three-state Potts CFT point, with chiral transitions (non-CFT) occurring below and above this point. Similarly, the $\mathbb{Z}_4^+$ and the $\mathbb{Z}_4$ phases can directly transition into the disordered phase via an AT CFT point, also flanked by chiral transition lines below and above. These chiral transition lines terminate at Lifshitz points and split into a BKT line and a PT line, which enclose the floating phase. The floating phase is thus separated from the disordered phase by a BKT transition and from the crystalline orders by a PT transition. Notice that both the chiral and PT transitions are commensurate-to-incommensurate: the former separates commensurate ordered regimes from incommensurate disordered ones, while the latter separates them from the incommensurate gapless floating phase. In the following, we focus on the direct transitions between the disordered phase and the crystalline orders and analyze their critical properties.

\subsection{The disordered phase}
\label{subsec:disorder}
\begin{figure}[t]
    \centering
    \includegraphics[width=1\linewidth]{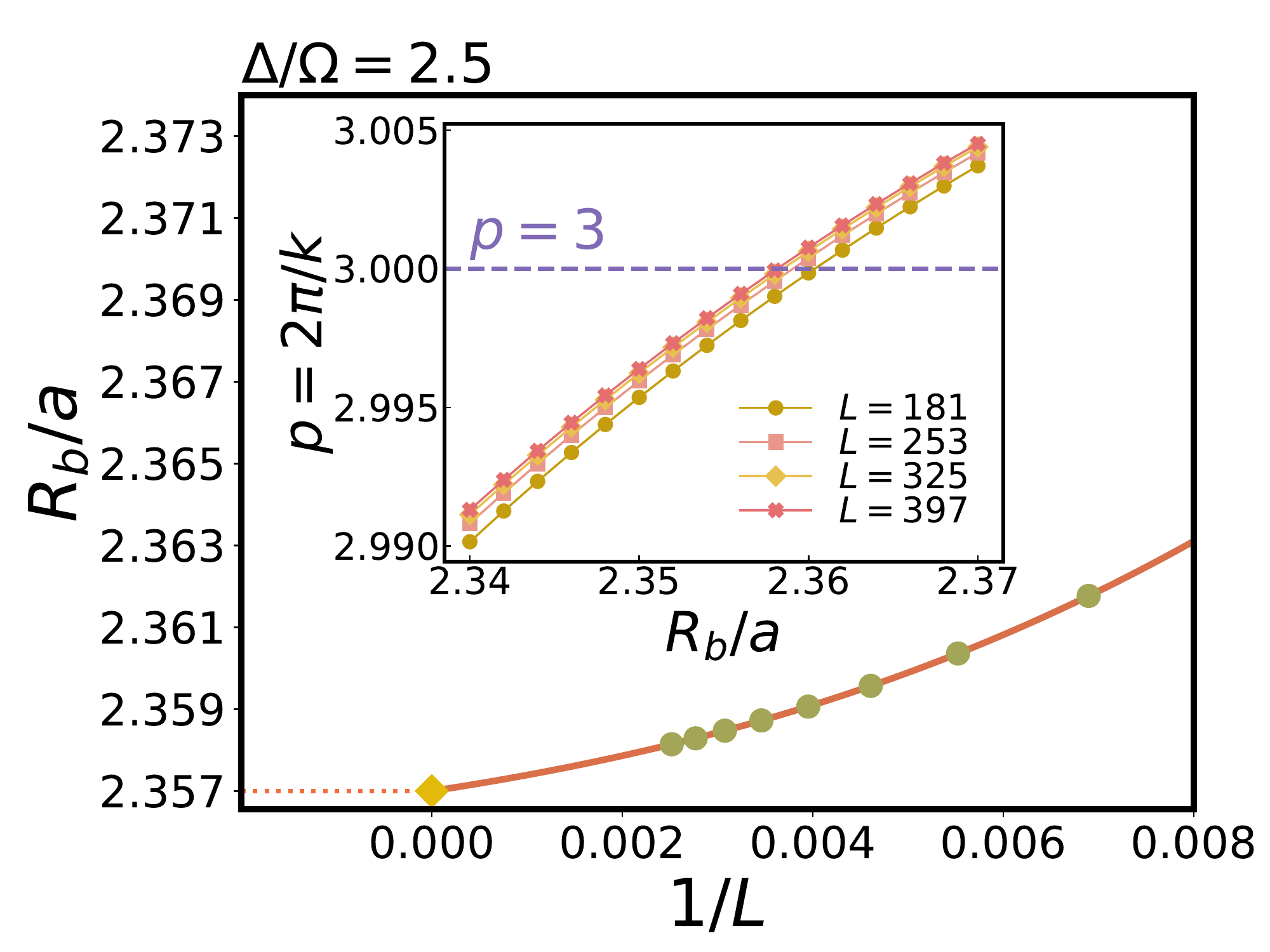}
    \caption{Extrapolation procedure to determine parameters with commensurate wave vector in the thermodynamic limit. The results are for $p = 3$ along the $\Delta/\Omega = 2.5$ cut in the disordered phase. The inset shows the peak position of the structure factor $S_B(k)$ as a function of $R_b/a$ for different system sizes, with data points connected by spline interpolations. The values of $R_b/a$ with $p = 3$ for each $L$ are determined and plotted as a function of $1/L$ in the main figure. The data is fit to a high-degree polynomial and extrapolated to $L \to \infty$ (yellow diamond).}
    \label{fig:ClineExtrapolation}
\end{figure}

Although the disordered phase is featureless at long range, its correlation function exhibits short-range oscillations with a wave vector that varies continuously with system parameters. As $R_b/a$ increases, the repulsive interaction strengthens, leading to an increase in the oscillation wavelength. Near the critical lines, the correlation length in the disordered phase diverges, and the oscillations of the correlation function become (quasi-)long-range. As shown analytically in Appendix~\ref{apsec:strucpeak}, the structure factor’s peak accurately determines the wavelength of even short-range density waves. We use the structure factor of bond-order correlations in Eq.~\eqref{eq:strucb} to extract the wave vector $k_o$, with the procedure illustrated in Fig.~\ref{fig:DisorderStrucFit}(a). Specifically, $S_B(k)$ is computed for finely spaced $k$ values, and spline interpolation is applied to identify the peak positions. Four typical cases of $k_o$ around $2\pi/3$ and $2\pi/4$ are presented. To verify the accuracy of this method, we also fit the correlation function $C_B(i, r)$ to the Ornstein-Zernike form $\cos(k_o r+\phi_0) \exp(-r/\xi) / \sqrt{r}$ \cite{OrnsteinCorr} and compare the resulting $k_o$ values from both methods. As shown in Fig.~\ref{fig:DisorderStrucFit}(b), the difference between the two methods is negligible and decreases rapidly as the correlation length increases and the wave vector approaches $\pi/2$, consistent with the derivation in Appendix~\ref{apsec:strucpeak}.

We use $S_B(k)$ to map the constant-$p$ lines in the disordered phase, as shown in Fig.~\ref{fig:phasediagram}(a), confirming that the wave vector (and wavelength) varies continuously with system parameters. As $\Delta/\Omega$ increases, incommensurate lines with non-integer $p$ enter the floating phase via BKT transitions. Commensurate lines with integer $p$ are located between the incommensurate lines where $p$ approaches integer values, as indicated in Fig.~\ref{fig:phasediagram}(a). These commensurate lines touch the lobes of the crystalline orders at CFT points. Locating these CFT points requires determining the commensurate lines in the thermodynamic limit. In Fig.~\ref{fig:ClineExtrapolation}, we outline the extrapolation procedure for identifying the parameters where the wavelength $p = 3$ in the disordered phase. For $\Delta/\Omega = 2.5$, the values of $p$ are calculated for various $R_b/a$ values across different system sizes $L$. Using spline interpolation, we determine the $R_b/a$ values corresponding to $p = 3$, denoted as $(R_b/a)_{p=3}$, for each $L$, as shown in the inset of Fig.~\ref{fig:ClineExtrapolation}. Finally, we fit $(R_b/a)_{p=3}$ as a function of $1/L$ to a high-degree polynomial to estimate $(R_b/a)_{p=3}$ in the thermodynamic limit ($L \to \infty$). This same procedure is applied to locate the commensurate line with $p = 4$ in the thermodynamic limit. There is a commensurate regime with $p = 2$ below the $p = 2$ line in Fig.~\ref{fig:phasediagram}(a), thus the boundaries of the $\mathbb{Z}_2^+$ and $\mathbb{Z}_2$ phases consist of Ising CFT points.

\begin{figure}[t]
\centering    \includegraphics[width=1\linewidth]{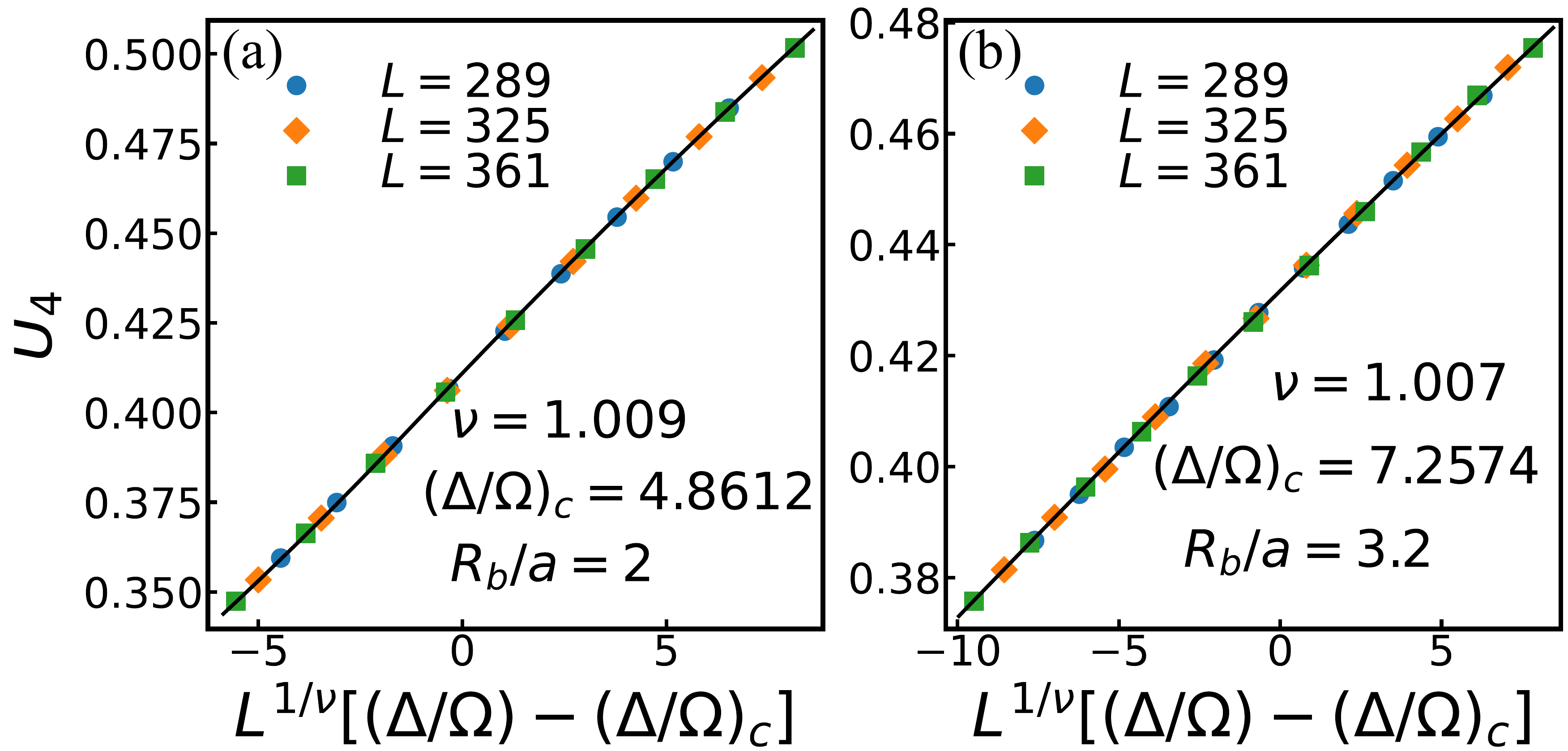}
\caption{Data collapse of the Binder cumulant for Ising phase transitions (a) between the $\mathbb{Z}_2^+$ order and the $\mathbb{Z}_4$ order on the $R_b/a=2$ cut, and (b) between the $\mathbb{Z}_3^+$ order and the $\mathbb{Z}_6$ order on the $R_b/a=3.2$ cut.}
\label{fig:ZppIsingBC}
\end{figure}

\subsection{Ising transitions}
\label{subsec:ising}
Both the transitions from the disordered phase to the $\mathbb{Z}_2^+$ and $\mathbb{Z}_2$ phases involve spontaneous translational $\mathbb{Z}_2$ symmetry breaking. The transition from the $\mathbb{Z}_2^+$ phase to the $\mathbb{Z}_2 \times \mathbb{Z}_2$ phase is associated with spontaneous top-bottom reflection $\mathbb{Z}_2$ symmetry breaking. The transition from the $\mathbb{Z}_2$ phase to the $\mathbb{Z}_2 \times \mathbb{Z}_2$ phase also exhibits translational $\mathbb{Z}_2$ symmetry breaking if the sites are numbered in a snake-like order starting from the top-left site, as shown in Fig.~\ref{fig:phasediagram}(c). Consequently, all these transitions belong to the Ising universality class, with numerical evidence provided in Ref.~\cite{JinPRDCritical2024}.

\begin{figure*}[ht]
\centering 
\includegraphics[width=1\textwidth]{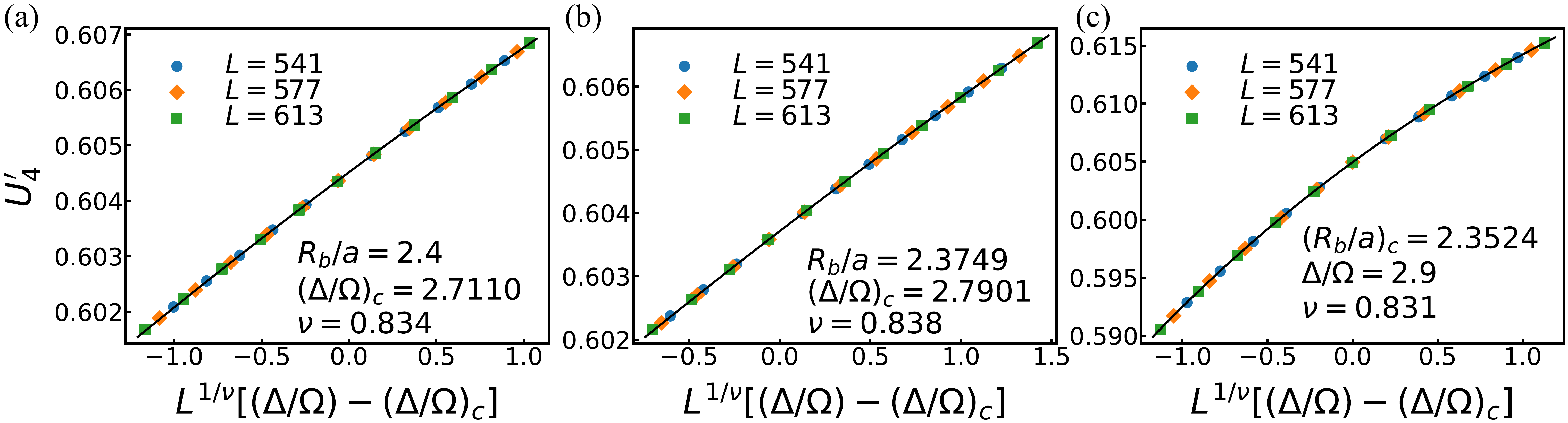}
\caption{Data collapse of the Binder cumulant for quantum phase transitions between the $\mathbb{Z}_3^+$ order and the disordered phase: (a) along the $R_b/a = 2.4$ cut, (b) at the CFT point along the $R_b/a = 2.3749$ cut, and (c) along the $\Delta/\Omega = 2.9$ cut. Solid lines represent polynomial fits to the collapsed data points.}
\label{fig:U4ofZ3+datacollapse}
\end{figure*}

Within the $\mathbb{Z}_2^+$ and $\mathbb{Z}_3^+$ lobes, spontaneous top-bottom reflection symmetry breaking can lead to transitions into the $\mathbb{Z}_4$ and $\mathbb{Z}_6$ phases, respectively. These transitions also fall under the Ising universality class. In Fig.~\ref{fig:ZppIsingBC}, we calculate the Binder cumulant $U_4$ for the order parameters $\hat{M}_{k=2\pi/4}^m$ and $\hat{M}_{k=2\pi/6}^m$ near these transitions and perform data collapses to determine the critical points and critical exponent $\nu$. For the $\mathbb{Z}_4$ and $\mathbb{Z}_6$ phases, we examine the $R_b/a = 2$ and $R_b/a = 3.2$ cuts, respectively. Around the crossing points of $U_4$ for system sizes $L = 289, 325, 361$, we compute values of $U_4$ in a narrow $\Delta/\Omega$ window of width approximately $0.01$ with a step size of $0.001$. The values of $U_4$ are plotted as a function of $L^{1/\nu}(\Delta/\Omega - (\Delta/\Omega)_c)$ and fitted to an 8-degree polynomial. By optimizing $\nu$ and $(\Delta/\Omega)_c$, the best data collapses are obtained at $[\nu, (\Delta/\Omega)_c] = (1.009, 4.8612)$ for the $\mathbb{Z}_4$ phase and $(1.007, 7.2574)$ for the $\mathbb{Z}_6$ phase. These values of $\nu$ agree well with the Ising universality class prediction, $\nu = 1$.

\subsection{$\mathbb{Z}_{3}^{+}$ order to the disordered phase}
\label{subsec:z3pchiral}

As previously mentioned, crystalline orders can directly melt into the disordered phase via continuous transitions, which are commensurate-to-incommensurate and belong to the chiral universality class, except at one CFT point. This CFT point resides at the intersection of the $\mathbb{Z}_3^+$ phase boundary and the commensurate line determined in Sec.~\ref{subsec:disorder}. To accurately locate the $\mathbb{Z}_3^+$ phase boundary, we perform a data collapse analysis of the Binder cumulant for $\hat{M}^B_{k=2\pi/3}$.

The results for the $R_b/a=2.4$ cut are shown in Fig. \ref{fig:U4ofZ3+datacollapse}(a), where the Binder
cumulant $U_4$ is calculated for system sizes $L = 541, 577, 613$. A $1/L$ correction term is introduced, and the modified cumulant, $U^{\prime}_4 = U_4(1+b/L)$, is used for data collapse, with $b$ as a tuning parameter. The best collapse is found at $(\Delta/\Omega)_c = 2.7110$ and $\nu = 0.834$, where data from all system sizes collapse onto a single curve. Another example for the $\Delta/\Omega = 2.9$ cut is shown in Fig. \ref{fig:U4ofZ3+datacollapse}(c) with a best data collapse at $(R_b/a)_c = 2.3524$ and $\nu = 0.831$.

\begin{figure}[h]
    \centering
    \includegraphics[width=0.9\linewidth]{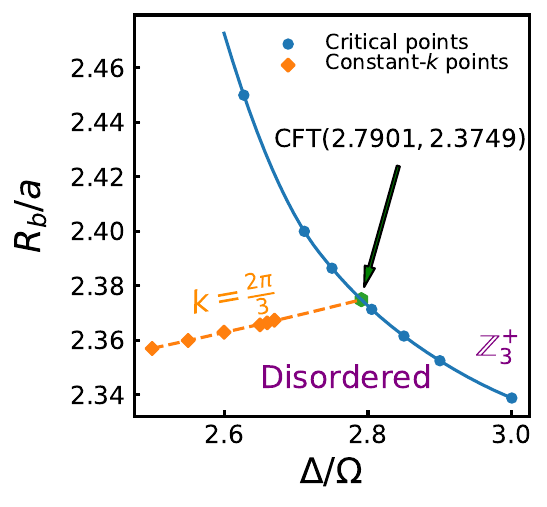}
    \caption{The phase boundary of the $\mathbb{Z}_3^+$ order and the commensurate line with $k = 2\pi/3$ in the disordered phase. Solid circles represent phase transition points determined from data collapse of the Binder cumulant, connected by spline interpolations. Diamonds indicate extrapolated parameter points with $k = 2\pi/3$ oscillations in the correlation function in the thermodynamic limit. The extrapolated diamonds determine the CFT point at the intersection of the two curves.}
    \label{fig:z3pcft}
\end{figure}

\begin{figure}[h]
    \centering
    \includegraphics[width=0.9\linewidth]{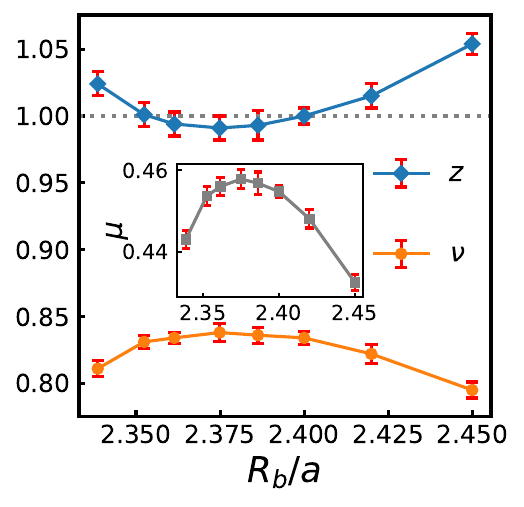}
    \caption{The correlation length exponent $\nu$ and the dynamical exponent $z$ for direct phase transitions between the $\mathbb{Z}_{3}^{+}$ order and the disordered phase. The inset displays the values of the Kibble-Zurek exponent, $\mu = \nu/(1+z\nu)$. The results focus on the vicinity of the CFT point.}
    \label{fig:criticalExponentofZ3+}
\end{figure}

Seven phase transition points on the $\mathbb{Z}_3^+$ lobe are located and plotted in Fig. \ref{fig:z3pcft}, connected by spline interpolation. The commensurate line is identified by extrapolating the points with a wave vector $k = 2\pi/3$ in the thermodynamic limit. The CFT point, located at $(\Delta/\Omega, R_b/a) = (2.7901, 2.3749)$, is determined where the commensurate line intersects the $\mathbb{Z}_3^+$ boundary. The Binder cumulant at the CFT point is shown in Fig. \ref{fig:z3pcft}(b) and the best data collapse is achieved at $\nu = 0.838$. This value is consistent with the three-state Potts CFT prediction of $\nu = 5/6$, with a minor discrepancy of $0.6\%$. The value of $\nu$ as a function of $R_b/a$ is plotted in Fig. \ref{fig:criticalExponentofZ3+}, showing that $\nu$ reaches its maximum value at the CFT point. This observation is consistent with results reported in Refs.~\cite{YuFidRyd2022, zhang2024probingquantumfloatingphases}.
\begin{figure}[h]
    \centering
    \includegraphics[width=0.9\linewidth]{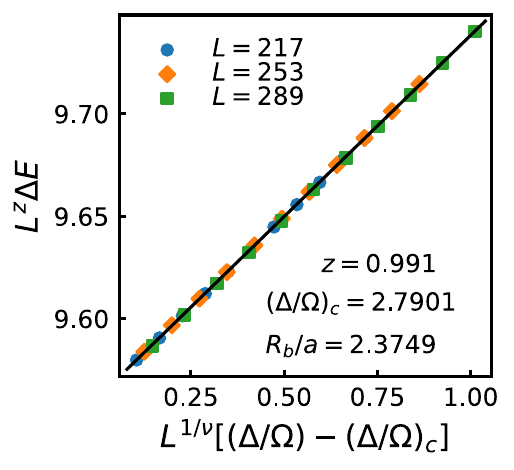}
    \caption{The data collapse of the rescaled energy gap $L^z\Delta E$ is performed for the CFT point on the $\mathbb{Z}_3^+$ boundary.}
    \label{fig:z3dEcft}
\end{figure}

For the parameters where $\nu$ is calculated, we also perform data collapse for the rescaled energy gap, $L^z \Delta E$, where $z$ is the dynamical exponent. An example of data collapse at the CFT point is shown in Fig.~\ref{fig:z3dEcft}, where energy gaps for $L = 217, 253, 289$ are calculated across a small range of $\Delta/\Omega$ values near the critical point. Using the $(\Delta/\Omega)_c$ and $\nu$ values obtained from the Binder cumulant, the best data collapse for the energy gap is achieved at $z = 0.991$, consistent with the CFT prediction of $z = 1$. The slightly smaller calculated value of $z$ can be attributed to finite-size effects and the finite bond dimension of the MPS in DMRG, which reduce the correlation length and increase the energy gap, requiring a smaller $z$ to satisfy the universal scaling form in Eq.~\eqref{eq:dEuniversalfunction}. The values of $z$ for other parameters are calculated similarly and plotted in Fig.~\ref{fig:criticalExponentofZ3+}, showing that $z$ reaches its minimum at the CFT point. The Kibble Zurek exponent $\mu = \nu/(1+z\nu)$ \cite{TWBKibble_1976,Zurek1985CosHelium,Polkovnikov2005dynamics,ZurekDorner2005,DziarmagaDynamicsIsing2005} is calculated and presented in the inset of Fig.~\ref{fig:criticalExponentofZ3+}. One can see that $\mu$ also reaches its maximum about $0.46$ at the CFT point, which can be tested experimentally in Rydberg-atom arrays \cite{Keesling2019Kibble}. Our results demonstrate that around the CFT point on the $\mathbb{Z}_3^+$ boundary, the transitions are non-CFT chiral transitions with $z > 1$. The value of $z$ for chiral transitions continuously increases, either above or below the CFT point along the phase boundary, until it reaches the Lifshitz point with $z = 2$, where the floating phase emerges. Beyond the Lifshitz point, the melting of the $\mathbb{Z}_3^+$ order is expected to proceed through an intermediate critical floating phase.

\subsection{$\mathbb{Z}_{4}^{+}$ order to the disordered phase}
\label{subsec:z4pchiral}

\begin{figure}[h]
\centering 
\includegraphics[width=0.9\linewidth]{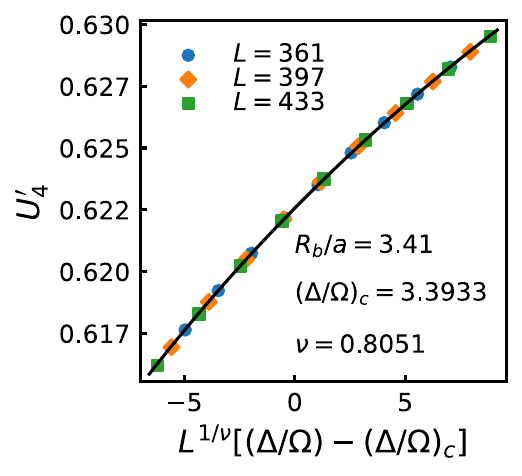}
\caption{The same as Fig.~\ref{fig:U4ofZ3+datacollapse}, but for the $R_b/a = 3.41$ cut.}
\label{fig:U4ofZ4+datacollapse}
\end{figure}

\begin{figure}[h]
    \centering
    \includegraphics[width=0.9\linewidth]{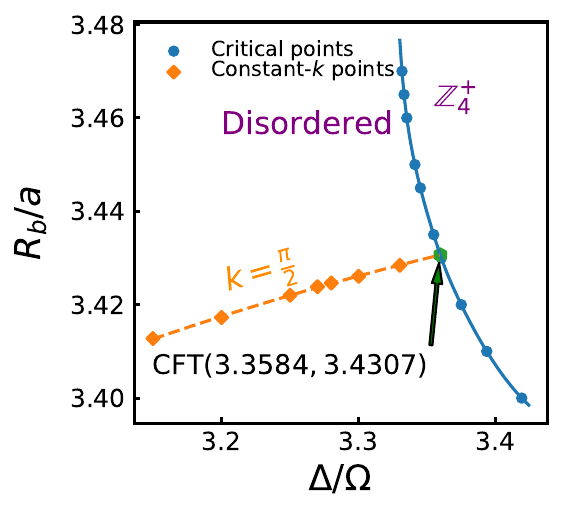}
    \caption{The same as Fig.~\ref{fig:z3pcft}, but for the $\mathbb{Z}_4^+$ order.}
    \label{fig:z4pcft}
\end{figure}

The melting process of the $\mathbb{Z}_4^+$ order is similar to that of the $\mathbb{Z}_3^+$ order. The direct transitions between the $\mathbb{Z}_4^+$ order and the disordered phase are predominantly chiral, except at a single CFT point located on the boundary of the $\mathbb{Z}_4^+$ lobe in Fig.~\ref{fig:phasediagram}(a). This CFT point belongs to the Ashkin-Teller universality class \cite{AshkinTeller1943}, where the correlation length exponent $\nu$ varies from $2/3$ (four-state Potts model) to $1$ (four-state clock model) \cite{Kohmoto1981AT}. For $0.683 < \nu \lesssim 0.82$, chiral perturbations are relevant, and the transitions become chiral immediately upon moving either up or down away from the CFT point. When $\nu \gtrsim 0.82$, the chiral perturbations are strong enough to cause the floating phase to emerge directly, either above or below the CFT point \cite{schulz1983phase,NyckeesPRR2022AT,LuscherATchiral2023}.

Figure~\ref{fig:U4ofZ4+datacollapse} illustrates the data collapse of the modified Binder cumulant $U^{\prime}_4$ for the order parameter $\hat{M}^B_{k = 2\pi/4}$ along the $R_b/a = 3.41$ cut, crossing the $\mathbb{Z}_4^+$ lobe boundary. The best data collapse is obtained at $(\Delta/\Omega)_c = 3.3933$ and $\nu = 0.8051$. Ten critical points on the $\mathbb{Z}_4^+$ lobe are determined using this approach, as shown in Fig.~\ref{fig:z4pcft}. Additionally, the parameters corresponding to $k = \pi/2$ in the thermodynamic limit are determined by extrapolation of the structure factor peak positions, also presented in Fig.~\ref{fig:z4pcft}. The data points are connected by spline interpolations, and the CFT point $(\Delta/\Omega, R_b/a) = (3.3584, 3.4307)$ is located at the intersection of the $\mathbb{Z}_4^+$ boundary and the commensurate line with $k = \pi/2$. Binder cumulant analysis gives $\nu = 0.8076$ for the CFT point, confirming that chiral perturbations are relevant, resulting in direct chiral transitions immediately away from the CFT point.

\begin{figure}[h]
    \centering
    \includegraphics[width=0.9\linewidth]{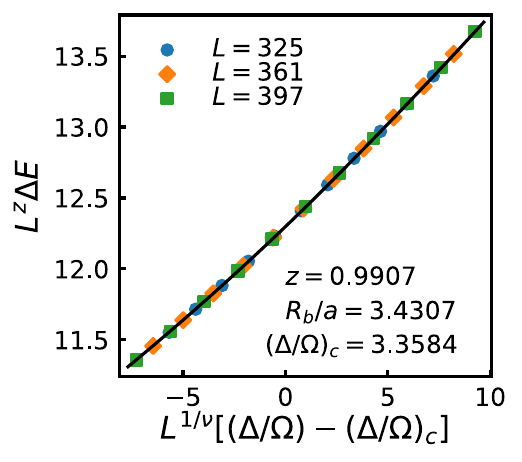}
    \caption{The same as Fig.~\ref{fig:z3dEcft}, but for the CFT point on the $\mathbb{Z}_4^+$ boundary.}
    \label{fig:z4dEcft}
\end{figure}

\begin{figure}[h]
    \centering
    \includegraphics[width=0.9\linewidth]{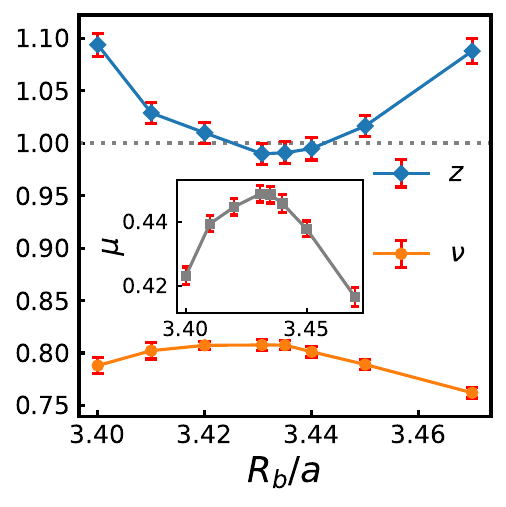}

    \caption{The same as Fig.~\ref{fig:criticalExponentofZ3+}, but for the $\mathbb{Z}_4^+$ order.}
    \label{fig:criticalExponentofZ4+}
\end{figure}

The data collapse of the energy gap for the CFT point is shown in Fig.~\ref{fig:z4dEcft}, where $z = 0.9907$ is obtained, consistent with the CFT prediction of $z = 1$. The exponents $\nu$ and $z$ are plotted in Fig.~\ref{fig:criticalExponentofZ4+}, showing that the CFT point corresponds to the maximum value of $\nu$ and the minimum value of $z$. Notably, the Kibble-Zurek exponent also reaches its maximum value of $\mu \approx 0.45$ at the CFT point. These features are the same as the behaviors observed for the $\mathbb{Z}_3^+$ phase.

\subsection{Floating phase} \label{subsec:floating}

\begin{figure*}[t]
\centering
\includegraphics[width=1\textwidth]{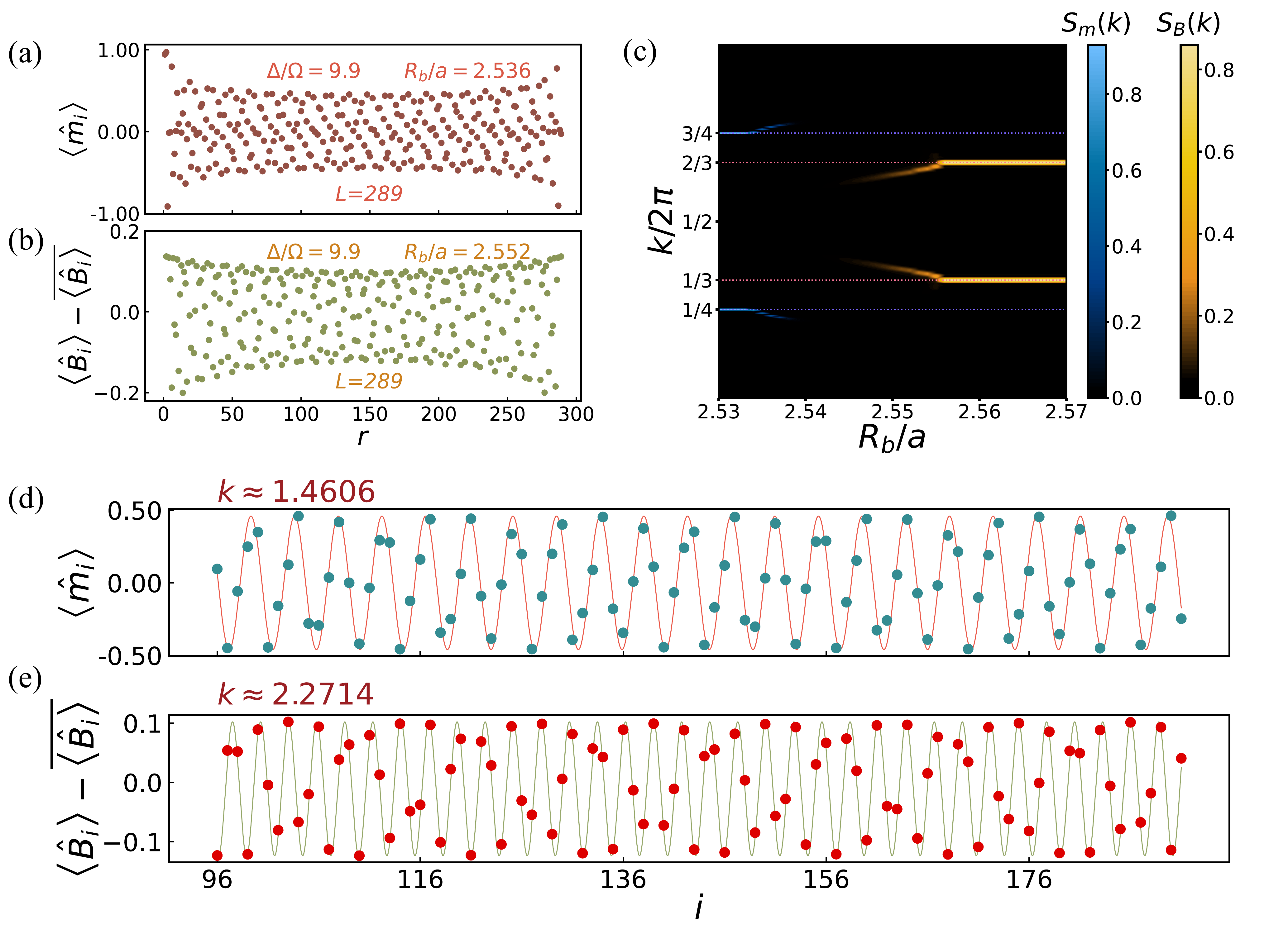}
\caption{(a) The profile of the density difference per rung, $\langle \hat{m}_{i}\rangle = \langle \hat{n}_{i,2} - \hat{n}_{i,1}\rangle$, in the floating phase labeled as FL. (b) The profile of bond orders, $\langle \hat{B}_{i}\rangle = \langle \hat{a}^\dagger_{i,1}\hat{a}_{i,2} + h.c.\rangle$, in the other floating phase labeled as FL$^{+}$. The average value of $\langle \hat{B}_{i}\rangle$ across all rungs is subtracted from the profile. (c) Structure factors $S_{m}(k)$ and $S_{B}(k)$ for $\langle \hat{m}_{i}\rangle$ and $\langle \hat{B}_{i}\rangle$, respectively, along the $\Delta/\Omega = 9.9$ cut. (d), (e) The middle segments of the profiles in (a) and (b) for the rung index range $i \in [96,191]$, fitted to the function $A\sin{(ki + \phi)}$. The extracted wave vectors are $k \approx 1.4606$ and $k \approx 2.2714$, respectively.
}
\label{fig:FloatingAlldensitymap&correlationmap}
\end{figure*}

Our phase diagram in Fig.~\ref{fig:phasediagram}(a) reveals the existence of two distinct floating phases located between the $\mathbb{Z}_4$ and $\mathbb{Z}_3^+$ phases. The floating phase adjacent to the $\mathbb{Z}_3^+$ phase, labeled FL$^+$, exhibits QLRO characterized by algebraically decaying bond-bond correlation functions $C_B(i,r)$. Conversely, the floating phase adjacent to the $\mathbb{Z}_4$ phase, labeled FL, shows QLRO described by algebraically decaying density-difference correlation functions $C_m(i,r)$. Between FL$^+$ and FL lies the disordered phase, where correlations decay exponentially. FL, FL$^+$, and the disordered phase are all part of the incommensurate regime separating two crystalline orders. In this regime, the wave vector of the oscillations in the correlation functions varies continuously with system parameters. As shown in Fig.~\ref{fig:phasediagram}(a), incommensurate constant-$p$ lines are distributed continuously between the crystalline orders.

To illustrate the properties of these floating phases, we present typical profiles of $\langle \hat{m}_i \rangle$ and $\langle \hat{B}_i \rangle$ for FL at $(\Delta/\Omega, R_b/a) = (9.9, 2.536)$ and FL$^+$ at $(\Delta/\Omega, R_b/a) = (9.9, 2.552)$, respectively, in Figs.~\ref{fig:FloatingAlldensitymap&correlationmap}(a) and (b). These results are obtained for a Rydberg ladder with $L=289$ rungs. Both $\langle \hat{m}_i \rangle$ and $\langle \hat{B}_i \rangle$ exhibit clear oscillations as the rung index $i$ varies, with the largest amplitudes near the edges that decay gradually toward the center. This behavior is consistent with the pattern of Friedel oscillations observed in critical phases of finite-size systems with open boundary conditions \cite{PhysRevResearch.3.023049,PhysRevB.108.054509}. Notably, no commensurate order signals are observed; instead, the oscillations exhibit incommensurate periodicities. To quantify this, we fit the middle portion of the profiles ($96 \leq i \leq 191$) to the trigonometric function $A\sin(ki+\phi)$, where $A$ is a constant. The fitted results, shown in Figs.~\ref{fig:FloatingAlldensitymap&correlationmap}(d) and (e), reveal wave vectors $k \approx 1.4606$ for FL and $k \approx 2.2714$ for FL$^+$. These values are incommensurate, with the former lying between $2\pi/5$ and $2\pi/4$ and the latter between $2\pi/4$ and $2\pi/3$.

The wave vector can also be extracted from the structure factors. Figure~\ref{fig:FloatingAlldensitymap&correlationmap}(b) displays $S_m(k)$ and $S_B(k)$ along the $\Delta/\Omega = 9.9$ cut. For $R_b/a \lesssim 2.533$, the system resides in the $\mathbb{Z}_4$ phase, characterized by true long-range order with a period-4 density wave. In this regime, $S_m(k)$ exhibits sharp peaks at $k = 2\pi/4$ and $k = 2\pi \times 3/4$. For $R_b/a \gtrsim 2.556$, the system is in the $\mathbb{Z}_3^+$ phase, where $S_B(k)$ peaks at $k = 2\pi/3$ and $k = 2\pi \times 2/3$. The peak heights of both $S_m(k)$ and $S_B(k)$ are close to $1$, consistent with predictions of the two crystalline orders in the large $\Delta/\Omega$ limit using the normalization factors in Eqs.~\eqref{eq:strucm} and \eqref{eq:strucb}. Between these two ordered phases, the peak positions of the structure factors deviate continuously from the commensurate values, becoming incommensurate. The prominent heights of the incommensurate peaks indicate the presence of floating phases, as slow algebraic decay in correlation functions corresponds to a similarly slow decay in the structure factor. In contrast, within the disordered phase, the structure factor decays as $\xi/L$, resulting in invisible peaks in Fig.~\ref{fig:FloatingAlldensitymap&correlationmap}(c). As the blockade radius increases, the average distance between Rydberg states grows, causing the wave vector in the incommensurate regime to decrease (or increase) depending on whether it lies between $0$ and $\pi$ (or $\pi$ and $2\pi$).

The presence of the two types of floating phases is natural, as each $\mathbb{Z}_p$ and $\mathbb{Z}_p^+$ ordered phase can melt into the disordered phase via an intermediate QLRO phase. This intermediate phase is expected to preserve short-distance features akin to the neighboring crystalline phases, characterized by similar correlation functions but with different wave vectors. While our phase diagram only identifies signals of FL$^+$ between the $\mathbb{Z}_6$ and $\mathbb{Z}_4^+$ phases, FL should also exist, especially in the larger $\Delta/\Omega$ regimes. Finally, the transitions from FL$^+$ and FL into the disordered phase between them occur via BKT transitions, whereas their transitions into nearby crystalline orders proceed via PT transitions. Numerical evidence for BKT and PT transitions in floating phases within ladder systems is provided in Ref.~\cite{zhang2024probingquantumfloatingphases}. Notice that for a Rydberg ladder with $a_y = 2a_x$, no $\mathbb{Z}_p^+$ orders are found at $R_b/a_x < 3.5$ and only the floating phase labeled FL exist in the phase diagram \cite{zhang2024probingquantumfloatingphases}.


\section{Conclusion} \label{sec:conclusion}
In this work, we explored the phase diagram and critical properties of a Rydberg atom ladder system with lattice spacings $a_x = 2a_y$. The phase diagram reveals various crystalline orders, which can be classified into two types: those with rung bond-order density wave order preserving top-bottom reflection symmetry, labeled $\mathbb{Z}_p^+$, and those with Rydberg density wave order in each leg, shifted to break top-bottom reflection symmetry, labeled $\mathbb{Z}_{2p}$. The $\mathbb{Z}_p^+$ phases break only the translational $\mathbb{Z}_p$ symmetry, whereas the $\mathbb{Z}_{2p}$ phases break both the translational $\mathbb{Z}_{2p}$ symmetry and the top-bottom reflection $\mathbb{Z}_2$ symmetry. Crystalline orders can melt directly into the disordered phase via either a CFT point or non-CFT chiral transitions above or below the CFT point. Farther from the CFT point, the melting process becomes a two-step transition involving an intermediate QLRO floating phase.

We investigated the critical properties of transitions between the $\mathbb{Z}_3^+$ and disordered phases, as well as between the $\mathbb{Z}_4^+$ and disordered phases. Using the Binder cumulant, we determined phase transition points and correlation length exponents $\nu$. The commensurate line in the thermodynamic limit was obtained by extrapolating finite-size peaks of the structure factors in the disordered phase, and the CFT point was identified as the intersection of the $\mathbb{Z}_p^+$ phase boundary and the commensurate line. Dynamical exponents $z$ were extracted from energy gap scaling. The critical exponents $\nu$ and $z$ at the CFT points on the $\mathbb{Z}_3^+$ and $\mathbb{Z}_4^+$ boundaries are consistent with the universality classes of the three-state Potts model and the four-state Ashkin-Teller model, respectively, both of which predict $z=1$. Away from the CFT point, $\nu$ decreases and $z$ increases along the chiral transition lines, consistent with previous studies. These parameter dependencies can be experimentally probed using Rydberg atom tweezer arrays, where the Kibble-Zurek exponent $\mu = \nu / (1 + z\nu)$ is measurable \cite{Keesling2019Kibble}.

Two QLRO floating phases were identified as intermediate states in the two-step melting of crystalline orders: FL$^+$, characterized by a QLR rung bond-order density wave adjacent to $\mathbb{Z}_3^+$ phases, and FL, characterized by a QLR rung density difference wave adjacent to $\mathbb{Z}_4$ phases. The short-range correlation functions of the floating phases resemble those of their neighboring crystalline orders but with different wave vectors. Experimental observations of ladder $\mathbb{Z}_p$ orders and the FL phase in a Rydberg ladder with different aspect ratio \cite{zhang2024probingquantumfloatingphases} suggest that the $\mathbb{Z}_p^+$ and FL$^+$ phases could be experimentally realized on similar platforms. This demonstrates the versatility of Rydberg quantum simulators for investigating novel quantum phases and critical phenomena.

Several open questions remain. For instance, where do the Ising transition line separating the $\mathbb{Z}_p^+$ and $\mathbb{Z}_{2p}$ orders and the $\mathbb{Z}_p^+$ phase boundary converge? A recent study \cite{garcia2024Rydladder} suggests that these lines merge into an Ashkin-Teller line for the $\mathbb{Z}_2^+$ and $\mathbb{Z}_4$ orders, but the scenario for $\mathbb{Z}_3^+$ and $\mathbb{Z}_6$ orders remains unresolved. Additionally, at what value of $\Delta/\Omega$ can the existence of the FL phase between the $\mathbb{Z}_4^+$ and $\mathbb{Z}_6$ orders be most easily detected? Furthermore, can the BKT transition lines between the FL$^+$ phase and the disordered phase, the PT transition lines between the $\mathbb{Z}_p^+$ phase and the FL$^+$ phase, and the Lifshitz points where the chiral transition lines terminate, be precisely located? These questions require further investigation and will be addressed in future studies. Finally, we note that the recent work in Ref.~\cite{garcia2024Rydladder} explored the phase diagram of a Rydberg ladder with $a_x = a_y$, where the same crystalline orders, with the exception of our $\mathbb{Z}_2 \times \mathbb{Z}_2$ phase, were observed but labeled differently.

\begin{acknowledgments}
We thank Yannick Meurice, James Corona, and Shan-Wen Tsai for help discussions. This work was supported in part by the National Natural Science Foundation of China under Grants No. 12304172 (J.Z.), No. 11874095 (L.-P.Y.), and No. 12347101, Chongqing Natural Science Foundation under Grant No. CSTB2023NSCQ-MSX0048 (J.Z.).
\end{acknowledgments}

\appendix
\section{Peak of the structure factor}
\label{apsec:strucpeak}
If the correlation function decays rapidly with distance, only the terms corresponding to the shortest distances significantly contribute to the summation in the structure factor’s definition. This can lead to a mismatch between the peak position of the structure factor and the wave vector of the oscillations. Such a phenomenon may occur in the disordered phase or in topological phases lacking local order, where correlation functions decay exponentially with a small correlation length, $\xi$. However, this discrepancy diminishes rapidly and vanishes as $\xi$ becomes large.

Considering an exponentially decaying oscillation $\exp(-|r|/\xi)\cos(qr)$, for $0 < k, q < \pi$, its Fourier transform $\sum_{r=0}^{\infty}\cos(k r)\cos(q r)\exp(-r/\xi)$ has a peak position $k_o$ given by:
\begin{widetext}
\begin{eqnarray}
\label{eq:peakofSk}
k_o = \begin{cases}
    0,  &   \text{if $\cos(q) > \frac{4}{\cosh(1/\xi)+\sqrt{\cosh^2(1/\xi)+8}} $} \\
    \pi,  &  \text{if $\cos(q) < -\frac{4}{\cosh(1/\xi)+\sqrt{\cosh^2(1/\xi)+8}}  $} \\
    \cos^{-1}\left[ \cosh(1/\xi) / \cos(q) -\tan(q) \sqrt{\cosh^2(1/\xi)-\cos^2(q)}\right].  & \text{otherwise}
\end{cases} 
\end{eqnarray}
\end{widetext}
When the correlation length is very small, the structure factor peaks at $0$ or $\pi$ for a broad range of wave vectors. This occurs at small interactions and far from the critical regime, which is not the focus of our study. On the other hand, even for a moderate correlation length, the width of the windows for $k_o = 0, \pi$ become very small. For example, if $\xi = 3$, $k_o = 0$ for $0< q < 0.19$ and $k_o = \pi$ for $\pi - 0.19 < q < \pi$. In fact, for large $\xi$, keeping the leading terms, the expression of the peak position becomes
\begin{eqnarray}
\label{eq:approxko}
k_o = \begin{cases}
    0,  &   \text{if $0 < q \lesssim \sqrt{\frac{7}{24}}\frac{1}{\xi}$} \\
    \pi,  &  \text{if $\pi - \sqrt{\frac{7}{24}}\frac{1}{\xi} \lesssim q < \pi$} \\
    q - \frac{\cos(q)}{8\sin^3(q)}\frac{1}{\xi^4},  & \text{if $\sqrt{\frac{7}{24}}\frac{1}{\xi} \lesssim q \lesssim \pi - \sqrt{\frac{7}{24}}\frac{1}{\xi}$}
    \end{cases}   
\end{eqnarray}
For incommensurability outside the small windows for $k_m = 0, \pi$, the peak position of the structure factor converges quickly to $q$ with a correction term that vanishes with $1/\xi^4$.  Therefore, the structure factor can accurately detect almost all the incommensurability for the disordered phase with moderately large correlation length.

For a quasi-long-range order with algebraically slow-decaying correlation functions, it can be numerically demonstrated that oscillations of the form $\cos(qr)/r^\eta$ with $\eta \leq 2$ have a Fourier peak precisely at $q$, meaning the summation $\sum_{r=1}^\infty \cos(kr)\cos(qr)/r^\eta$ peaks exactly at $k = q$. In fact, for $\eta = 2$, using the relation between the Polylogarithm functions and the Bernoulli polynomials
\begin{eqnarray}
\label{eq:polybernoulli}
\operatorname{Li}_n\left(e^{2 \pi i x}\right)+(-1)^n \operatorname{Li}_n\left(e^{-2 \pi i x}\right)=-\frac{(2 \pi i)^n}{n!} B_n(x),
\end{eqnarray}
where
\begin{eqnarray}
\mathrm{Li}_s(z)&=&\sum_{k=1}^{\infty} \frac{z^k}{k^s}=z+\frac{z^2}{2^s}+\frac{z^3}{3^s}+\cdots, \\
B_2(x)&=&x^2-x+\frac{1}{6}
\end{eqnarray}
the Fourier transform $\sum_{r=1}^\infty \cos(kr)\cos(qr)/r^2$ is proportional to $(k+q)^2-2\pi (k+q) +4(k-q)^2-4\pi|k-q| + 4\pi^2/3$, which takes the maximal value at $k=q$ for $0 < k, q < \pi$. Then for $\eta < 2$, the correlation function decays slower, so the Fourier peak also resides at $k=q$. 

%
    
\end{document}